          [2001/04/25 1.1 (PWD)]
\documentclass[a4paper,twocolumn]{esapub2005} 
\usepackage{times}
\usepackage{natbib}
\usepackage[dvips]{graphicx}

\title{Detection of periodic signatures in the solar power spectrum.   \ \ \ \ \ \ \  On the track of $\ell$=1 gravity modes}
\author{R. A. Garc\'\i a$^{1,2}$}
\author{S. Turck-Chi\`eze$^{1,2}$}
\author{S. J. Jim\'enez-Reyes$^{3,4}$}
\author{J. Ballot$^5$}
\author{P. L. Pall\'e$^3$}
\author{A. Eff-Darwich$^6$}
\author{S. Mathur$^{1,2}$}
\author{J. Provost$^7$}

\affil{$^1$DAPNIA/DSM/Service d'Astrophysique, CEA/Saclay, 91191 Gig-sur-Yvette Cedex, France}
\affil{$^2$Unit\'e mixte de recherche CEA-CNRS-Universit\'e Paris VII- UMR 7158, 91191 Gif-sur-Yvette Cedex, France}
\affil{$^3$Instituto de Astrof\'\i sica de Canarias, 38205, La Laguna, Tenerife, Spain}
\affil{$^4$School of Physics \& Astronomy, Univ. of Birmingham, Edgbaston, Birmingham B15 2TT, Uk}
\affil{$^5$Max-Planck-Institut f\"ur Astrophysik, Karl-Schwarzschild-Str. 1, 85748 Garching, Germany}
\affil{$^6$Departamento de Edafolog\'\i a y Geolog\'\i a, Universidad de La Laguna, Tenerife, Spain}
\affil{$^7$D\'epartement Cassiop\'ee, UMR CNRS 6202, Observatoire de la C\^ote d'Azur, BP 4229, 06304 Nice Cedex 4, France}

\newcommand{\etal}{{\em et~al.}}

\begin{document}

\keywords{Sun: Helioseismology, Sun: Interior, Sun: Oscillations}

\maketitle

\begin{abstract}

In the present work we show robust indications of the existence of g modes in the Sun using 10 years of GOLF data. The present analysis is based on the exploitation of the collective properties of the predicted low-frequency (25 to 140 $\mu$Hz) g modes: their asymptotic nature, which implies a quasi equidistant separation of their periods for a given angular degree ($\ell$). The Power Spectrum (PS) of the Power Spectrum Density (PSD), reveals a significant structure indicating the presence of features (peaks) in the PSD with near equidistant periods corresponding to $\ell$=1 modes in the range n=-4 to n=-26. The study of its statistical significance of this feature was fully undertaken and complemented with Monte Carlo simulations. This structure has a confidence level better than 99.86~$\%$ not to be due to pure noise. Furthermore, a detailed study of this structure suggests that the gravity modes have a much more complex structure than the one initially expected (line-widths, magnetic splittings...). Compared to the latest solar models, the obtained results tend to favor a solar core rotating significantly faster than the rest of the radiative zone.  In the framework of the Phoebus group, we have also applied the same methodology to other helioseismology instruments on board SoHO and ground based networks.
\end{abstract}

\section{Introduction}
\begin{figure}[!hb]
\centering
\rotatebox{90}{\includegraphics[width=.37\textwidth]{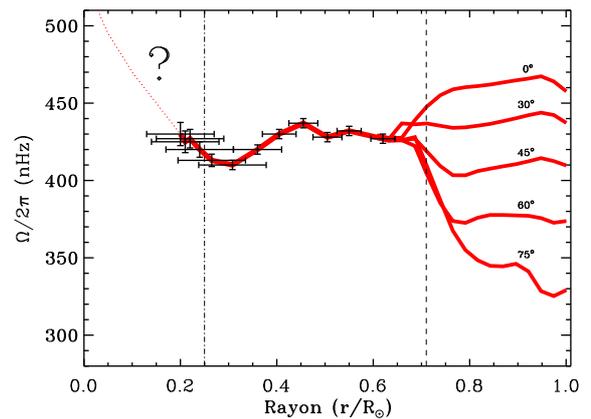}}
\caption{2-D inversion of the solar rotation profile using the modes $\ell \le 25$ from GOLF \citep{Garcia2004} and MDI \citep{korzennik2005} instruments. The horizontal error bars increase towards the centre of the Sun. Below 0.2 the rotation profile is unknown. }
\label{rota}
\end{figure}

\begin{figure}[!ht]
\centering
\rotatebox{90}{\includegraphics[width=.35\textwidth]{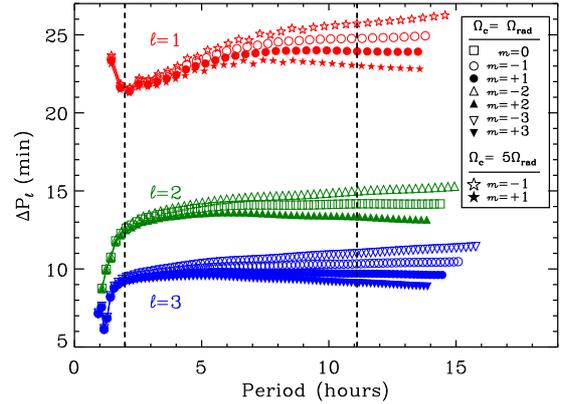}}
\caption{Separation in period of theoretical g modes with a rigid rotation rate in the core for $\ell$=1 (red), $\ell$=2 (green) and $\ell$=3 (blue) modes. The red stars correspond also to the same $\ell$=1 separations but for a model with a rotation rate in the core 5 times larger than in the rest of the radiative region. The region inside the vertical dashed lines corresponds to the region of interest (25 to 140 $\mu$Hz).}
\label{frec_teor}
\end{figure}

Detailed knowledge of the solar interior has been obtained by means of precise measurements of its acoustic (p) eigenmodes and the subsequent structural inversions, providing the stratification of crucial variables like the sound speed down to ~0.05 $R_\odot$. However, the acoustic modes are less sensitive to other structural variables such as the density. In this case, the agreement with the models is still poor in the deepest layers of the radiative zone \citep{STC2001}. Moreover the dynamical properties of the solar interior are practically unknown in the nuclear core. Indeed, due to the limitation in the number of acoustic modes penetrating this region and to their low sensitivity there, the solar rotation profile (see Fig.~\ref{rota}) is very uncertain below 0.2 $R_\odot$ \citep[e.g.][]{Couvidat2003, Garcia2004, Thompson2003}. At these depths, large uncertainties also exist because of their lack of sensitivity. On the contrary, solar gravity (g) modes exclusively provide information on the radiative region and the inner core. They are required to obtain a complete picture of the interior of our Sun.
Unfortunately, solar g modes have not yet been clearly detected although extensive and intensive searches have been developed along past decades. These modes are evanescent in the convective zone and their amplitudes at the solar surface could be very small \citep{Andersen1996,  Kumar1996}.  To look for the g modes, two complementary ways of searching strategies have been developed during the last decades:

\begin{itemize}
\item The first method consists in looking for individual modes (relevant spikes in the Fourier spectrum of the observed signal) above a given statistical threshold (typically 90 $\%$ confidence level) at frequencies above 150 $\mu$Hz. The first studies from the GOLF (Global Oscillations at Low Frequency) Team \citep{Gabriel2002} and the Phoebus \citep{App2000} and BiSON (Birmingham Solar Oscillations Network, \citep{Chaplin2002}) groups looked for individual peaks (spikes). An upper limit of 1~cm/s has been found for these single peaks. Another study of the GOLF data was devoted to the search for multiplets -- instead of individual peaks -- which has lowered the detection level to more than a half. Two candidates have been observed and followed with time with more than 90\% confidence level in a 20 $\mu$Hz interval \citep{STC2004} and one of them has even been detected with a 98\% confidence level after ~3000 days of observations in a 10 $\mu$Hz interval \citep{STC2005}. However, some ambiguity remains on the proper labelling of the detected components, in term of their $\ell$, $m$ and $n$ values. Nevertheless, some scenarios have been proposed in order to explain the visible peaks which can constrain the physics and dynamics of the solar core. Indeed, the frequencies of such modes are extremely sensitive to the structural properties of the radiative zone, including the problem of the solar abundances \citep{Mathur2006}.

\item An alternative way, developed in the early 80s, is to look for the g-mode asymptotic properties (see for example \citep{Palle1991, Hill1991} and references there in): this theory predicts a constant separation (in period) between the central components of the gravity modes at low frequency. Fig.~\ref{frec_teor} shows the separation in period of consecutive g modes ($\Delta P_{\ell}=P_{\ell,(n+1)}-P_{\ell,n}$), with $\ell \le 3$ and periods below 11 hours, as a function of the period computed from a seismic solar model \citep{Couvidat2003} assuming a rigid core rotation \citep{Provost2000}. The effect of the rotation is to split the modes into different $m$-components regularly spaced in frequency. As a consequence,  only zonal ones, $m$=0, are equidistant in period while the others shift progressively as the period of the modes increases. Consequently, this method becomes extremely sensitive to the rotation rate of the inner solar layers.

\end{itemize}

\begin{figure*}[ht!]
\centering
\begin{tabular}{@{}c@{}c@{}}
\rotatebox{90}{\includegraphics[width=.365\textwidth]{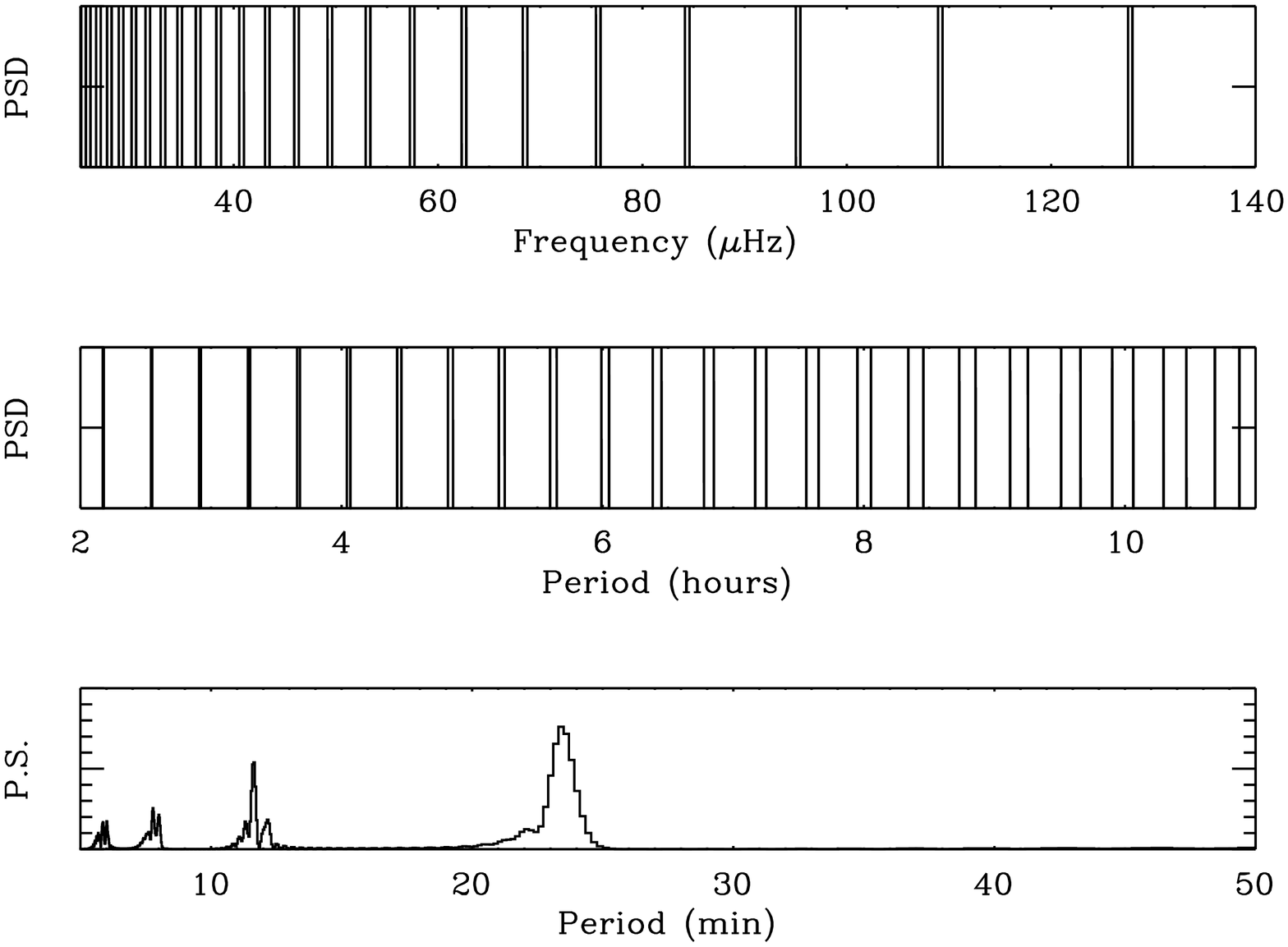}}&%
\rotatebox{90}{\includegraphics[width=.365\textwidth]{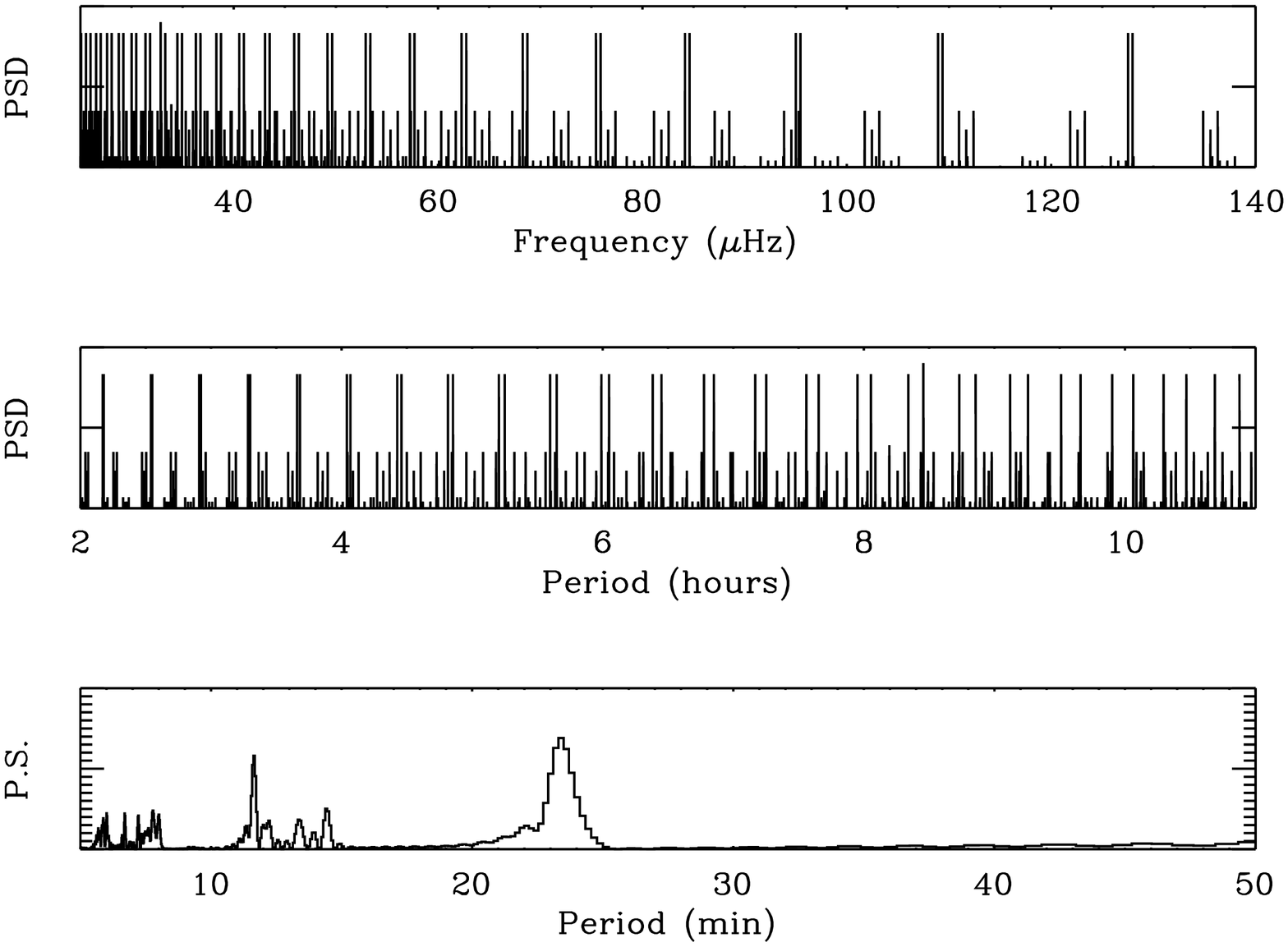}}\\
\end{tabular}
\caption{Simulated PSD of theoretical g modes computed from the seismic model using a rigid core rotation law for $\ell$=1 modes (left panels) and $1 \le \ell \le 3$ (right panels) expressed in frequency (top) and in period (middle). The bottom panels correspond to the PS of each above spectrum. In the bottom left panel a clear peak appears centered at 23.5 minutes corresponding to the main periodicity of the $\ell =1$ modes as well as all the harmonics at lower periods. In the bottom right panel the structure of the main separation of the $\ell$=2 modes appears centered at 14 minutes.}\label{modelos}
\end{figure*}

\section{Analysis technique}
Using the g modes showed in Fig.~\ref{frec_teor} and considering that they have long enough (more than ten years) lifetime and a constant amplitude for each mode degree (with the same relative visibility ratio than for the p modes), we can build up a simple model -- without noise in a first step-- of the power spectrum between 25 and 140 $\mu$Hz, $PSD(\nu$) (see top panels of  Fig.~\ref{modelos}). To analyze the existing periodicities inside this spectral region, the independent variable of $PSD(\nu)$ is changed from frequency to period, giving $PSD(P)$ (middle panels of Fig.~\ref{modelos}). Then, the periodogram of the $PSD(P)$, hereafter called PS, is computed using a 10-time-oversampled sine-wave fitting algorithm (bottom panels of Fig.~\ref{modelos}). The set of modes corresponding to each degree gives rise to a peak on the PS and to a sequence of harmonics of it. For the $\ell$=1 modes the main peak in the PS is centered at $\Delta P_1 \sim 23.5$ minutes with an asymmetric shape because the separation of the modes at lower periods decreases (see also Fig.~\ref{frec_teor}). The behavior of the $\ell$=2 and 3 modes can be explained in the same way. Looking at the PS including all the modes together (right bottom panel of Fig.~\ref{modelos}), only the $\Delta P_1$ can be easily discerned. The others, $\Delta P_{2,3}$,  are mixed with the harmonics of the $\ell$=1 and it is not easy to recover them. The presence of the $\ell$=2 modes have also an influence on the $\Delta P_1$ due to the beating between them. The contribution of the $\ell=3$ modes is negligible due to their small simulated visibility. For  these reasons we have decided to concentrate our study only on the signature of the $\Delta P_1$  between 22 and 26 minutes. It is also important to notice that a change in the relative amplitude between the different $\ell$ modes will modify the relative amplitudes of the peaks of each $\Delta P_{\ell}$.

\begin{figure}[!ht]
\centering
\includegraphics[width=.49\textwidth]{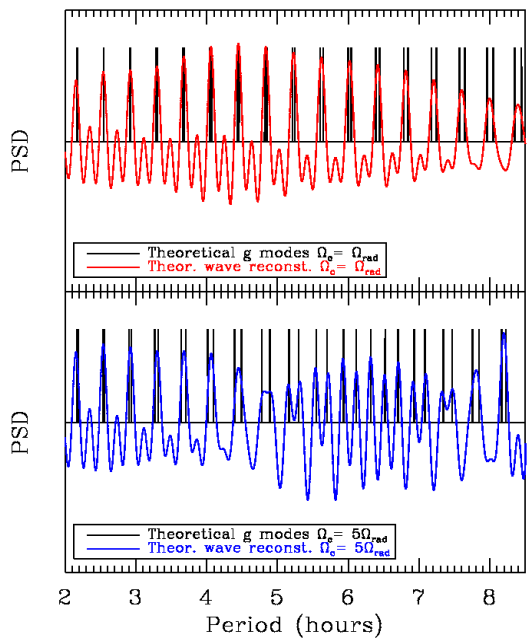}
\caption{Top: Zoom between 2 and 8.5 hours of the reconstructed waves in the PSD(P) using the g modes computed from the seismic model with a rigid core rotation. Bottom:Same as before but using a core rotation rate 5 time faster than the rest of the radiative region at $R=0.15~R_\odot$.}
\label{reco_th}
\end{figure}

From the $\Delta P_1$ pattern measured in the PS, it is possible to reconstruct the waves in the PSD(P) corresponding to the peaks between 22 and 26 minutes and its first harmonic. Thus, we compute:
\begin{equation}
\sum_{P_i\in[22;26]} A_i \sin\left(\frac{2\pi t}{P_i} + \phi_i\right)
+\sum_{P_i\in[11;13]} A_i \sin\left(\frac{2\pi t}{P_i} + \phi_i\right)	 \nonumber
\end{equation}
By doing so, we are also taking into account the phase of the waves in the PS. This information constrains the position of the maximum of the sinusoidal waves that should coincide with the positions of the g modes in the PSD. Fig.~\ref{reco_th} shows the reconstructed waves superimposed on the original $\ell$=1 simulated modes for the same solar seismic model but with two different core rotation laws. Firstly, the sinusoidal waves reproduce the pattern of the g modes and they are very sensitive to the splitting of the modes (see the bottom panel) where the waves are divided in two at ~5.5 hours to follow both split components. Thus, this method can be directly used to check the general trend of the solar rotation profile inside the core.

\section{Data analysis and discussion}

\begin{figure}[!ht]
\centering
\includegraphics[width=.47\textwidth]{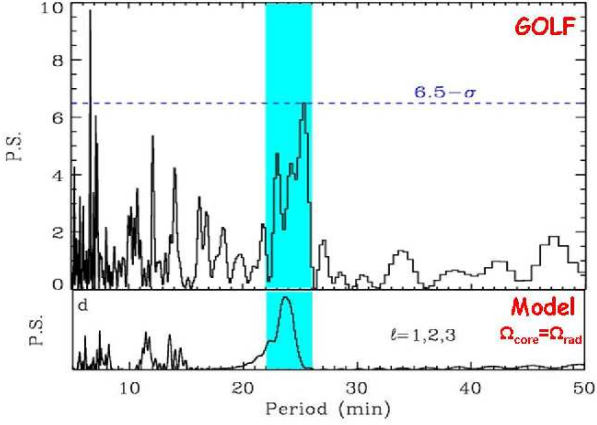}
\caption{Top: PS of the PSD expressed in period computed from the GOLF velocity time series. Bottom: PS of the $PSD(P)$ from theoretical g-modes computed from the seismic model and using a rigid core rotation.}
\label{PS_GOLF}
\end{figure}

Data from the GOLF instrument aboard the SoHO mission are used in this work. GOLF is a resonant scattering spectrophotometer measuring the line-of-sight Doppler velocity of the sodium doublet \citep{Gabriel1995}. We use 10 years time-series starting 1996 April 11 and properly calibrated into velocity following \citep{Garcia2005}. After computing the corresponding Fourier Transform, the solar background was properly modeled and removed by means of a first order polynomial fit in a log-log scale. The resultant $PSD(\nu)$ is flat in the region of interest and its statistical properties are well represented by a $\chi^2$ with 2 degrees of freedom (d.o.f.). The resultant PS is plotted in the top panel of Fig.~\ref{PS_GOLF}. Between 22 and 26 minutes a wide pattern of high amplitude peaks appears. The maximum amplitude in this region reaches 6.5-$\sigma$ while the average power is 2.95 times higher than the averaged power of the rest of the spectrum. Comparing this pattern with the one produced by the theoretical $\ell$=1 modes issued from the seismic model (bottom panel of Fig.~\ref{PS_GOLF}) we find that the PS are both alike.

\begin{figure}[!ht]
\centering
\includegraphics[width=.47\textwidth]{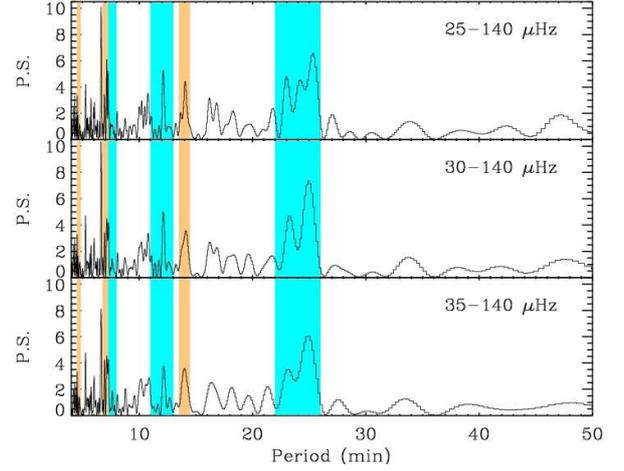}
\caption{PS of three different regions of the $PSD$, from top to bottom 25, 30 and 35 $\mu$Hz to 140 $\mu$Hz respectively. The cyan shaded zones correspond to where the $\Delta P_1$ structure is expected, between 22 to 26 min, and its first two harmonics. The orange shaded zones are the same but for the $m=0$ component of the $\Delta P_2$. }
\label{ventana}
\end{figure}

We have studied the $PSD(\nu)$ between 25 and 140 $\mu$Hz, because this region is a compromise between the need of having modes in the asymptotic region and with high enough amplitudes (at lower periods, see: \citep{Kumar1996, Provost2000}). We have verified that when shorter spectral regions are used in the calculation of the PS, e.g., from 30 or 35 to 140 $\mu$Hz, the prominent structure around 22 to 26 minutes still appears in the PS  with maxima at 7.3- and 6-$\sigma$ respectively (see the three panels of 
Fig.~\ref{ventana}).

To check the probability that this kind of structure placed between 22 and 26 minutes (guided research) is due to pure noise ($H_o$ hypothesis test) we compute a Monte Carlo simulation of 6 $10^5$ realizations (uncertainty on the results of 0.13 $\%$) of a pure noise power spectrum following a $\chi^2$ with 2 d.o.f. \citep{Garcia2006}. To characterize this feature, we use two indicators: the maximum power in the region between 22-26 minutes (6.5-$\sigma$) and the amount of total power in this region (2.95 above the rest of the spectrum). Thus, with 99.49 $\%$ of confidence level, we reject the $H_0$ hypothesis. In the framework of the PHOEBUS group \citep{App2006} we also compute the probability to find this kind of pattern but anywhere in the PS (non guided research) between 5 and 50 minutes. In this case, with a confidence level of 94.15$\%$, the $H_o$ hypothesis is rejected.  

\begin{figure}[!ht]
\centering
\includegraphics[width=.49\textwidth]{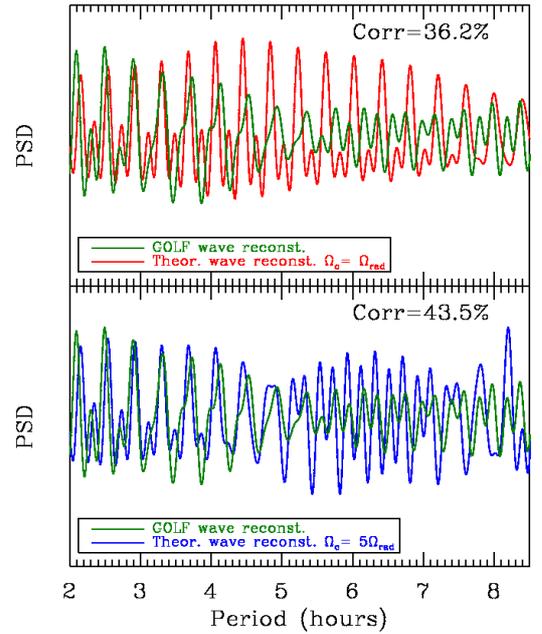}
\caption{Reconstructed waves using GOLF data compared to the reconstructed waves from the g modes of the seismic model using a rigid core rotation law (top) or using a 5 times faster core rotation law (bottom).}
\label{reco_gf}
\end{figure}

The Monte Carlo simulations show that  pure noise is very unlikely to generate this kind of patterns, then we can try to check if they can be due to periodic signals in the PSD placed where we expect the gravity modes to be. Therefore, we reconstruct the waves in the PSD(P) corresponding to the $\Delta P_1$ and its first harmonic. The result is plotted in Fig.~\ref{reco_gf}. To characterize it we compute the correlation of the resultant waves with the ones computed from solar models changing several parameters: the solar model, the rotation law and the solar core rotation axis inclination. In most of the cases we see that, for a given solar model, the correlation increases with a higher rotation rate. It reaches a maximum between 3 to 5 times the rotation of the rest of the radiative region, depending on the other parameters taken into account. This can be seen by  comparing both panels of Fig.~\ref{reco_gf}. In the top one, the GOLF reconstructed waves start to be larger at ~3.8 hours and are clearly split in two different waves at 5.5 hours while the model with a rigid rotation has single thinner peaks all over the plotted region. In the bottom panel the correlation between both curves is higher and their general behaviours are similar. Only above ~7.5 hours the GOLF waves shift to higher periods in comparison to the simulated ones.

Using the same Monte Carlo simulation as before, we can check again the $H_0$ hypothesis comparing the resultant reconstructed waves with a model an imposing a threshold of only 20$\%$ which is the smallest one we found when comparing the real GOLF data with the models. In this case $H_0$ is rejected with more than 99.86$\%$ confidence level.

\begin{figure}[!ht]
\centering
\includegraphics[width=.49\textwidth]{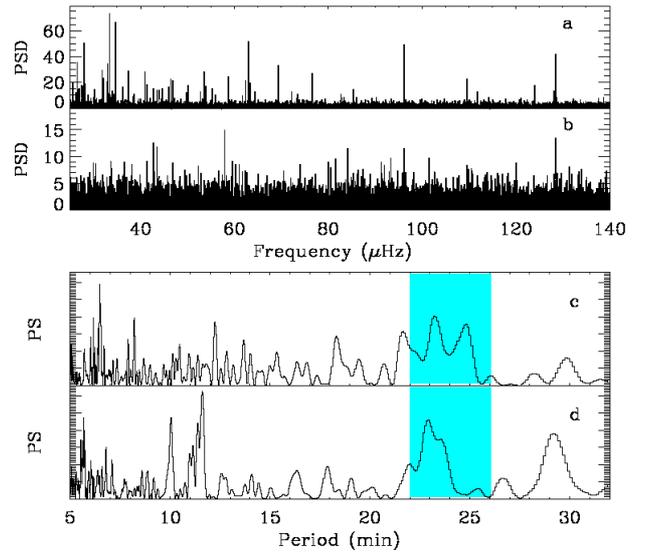}
\caption{PSD of two different simulations (a, b) and their corresponding PS spectra (c, d). See the text for more details.}
\label{ruido}
\end{figure}

Up to now, we have seen that the studied structure is unlikely due to pure noise. One could think that it may be due to some kind of instrumental noise. However there are no known instrumental effects having periodicities in the range of 2-14 hours and being equidistant in period. The other possibility is that these periodicities have a solar origin. Excepting a possible relation with convection we can analyze the consequences of assuming that they are produced by g-mode asymptotic properties. With such objective, we have introduced noise -- a $\chi^2$ with 2 d.o.f. -- to the simulations used previously. The resultant PSD (in arbitrary units) and corresponding PS are shown in Fig.~\ref{ruido}a and Fig.~\ref{ruido}c respectively (for a simulation with a Signal-to-Noise ratio (S/N) of 30. The visibility of the $\Delta P_1$ structure is degraded because of the noise. Instead of having a big peak -- like in the noise-free case -- we have a more complicated structure (the exact distribution of peaks depends on each noise realization). However, in order to keep a significant $\Delta P_1$ structure (above the limits defined in the Monte Carlo simulations) we need a very high signal/noise (S/N) ratio in the power spectrum, much higher than the GOLF observed one which seems to be around one. Otherwise we should have already measured the g modes.   

Therefore, it is not possible to find, in average, a proper $\Delta P_1$ -- with a small S/N ratio in the power spectra -- unless the probability to have energy on the modes is transfered into more bins. We have found three ways to reach such objective: a) assuming a finite lifetime of the modes as suggested by the recent works by \citep{Rogers2005} which spreads the power of each mode from a single bin into a Lorentzian profile covering a larger region; b) increasing the visible multiplet components assuming a different rotation axis inclination of the core than in the radiative region, as it was first suggested by \citep{Sturrock1990}; c) considering a magnetic field in the core or at the base of the convective zone. Indeed, \citep{Goode1992} and more recent calculations \citep{STC2005} have given a complicated picture of the contribution of a deep magnetic field with many components appearing for each mode. To explore these possibilities we have chosen the first. Thus, Fig.~\ref{ruido}b corresponds to the power spectrum of the same simulation as in Fig.~\ref{ruido}a but using a finite lifetime -- several months -- of the g modes. In this case the S/N can be significantly reduced (S/N=5) while the PS continue to show a detectable $\Delta P_1$ structure (See Fig.~\ref{ruido}d). To go further, a more statistical approach of these simulations is needed including heavy Monte Carlo simulations covering the different physical mechanisms explained above.  

\begin{figure*}[ht!]
\centering
\begin{tabular}{@{}c@{}c@{}}
\rotatebox{90}{\includegraphics[width=.365\textwidth]{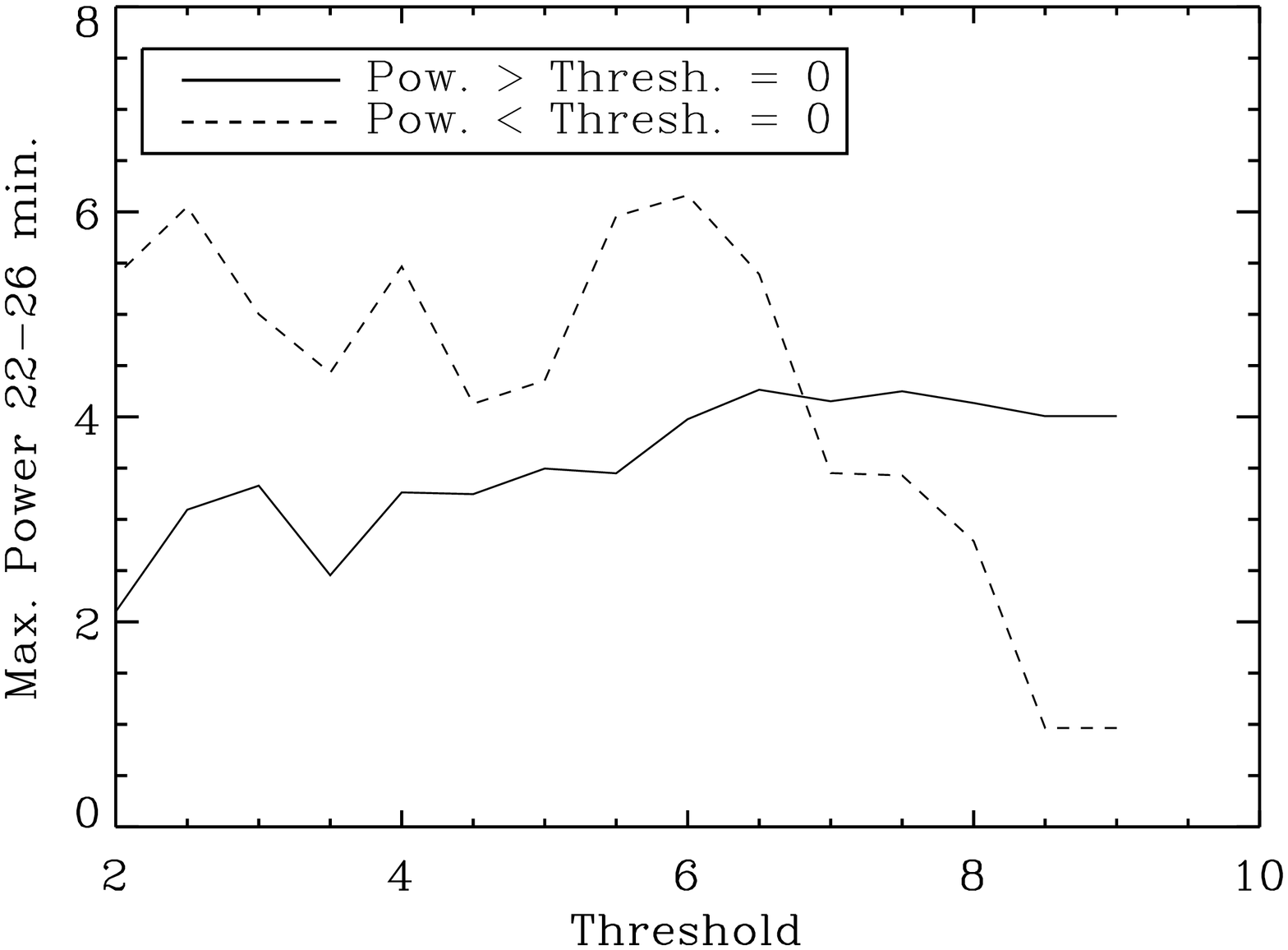}}&%
\rotatebox{90}{\includegraphics[width=.365\textwidth]{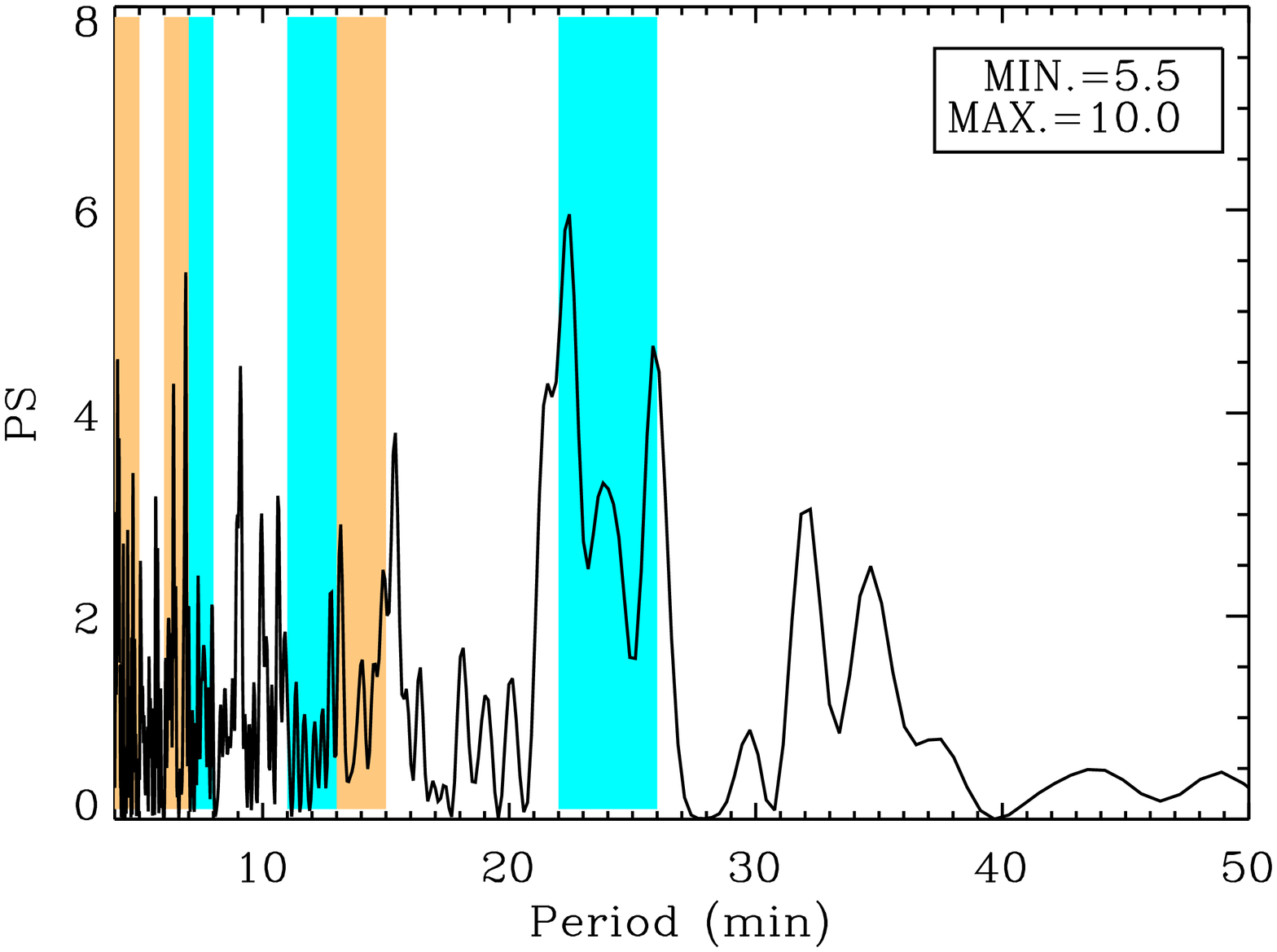}}\\
\end{tabular}
\caption{Left: Power of the highest peak between 22 and 26 minutes in the PS when the VIRGO/SPM PSD is thresholded above (continuous line) or below (dashed line) a given value between 2 and 9. Right: PS corresponding of the PSD computed using 3473 VIRGO/SPM intensity data. A minimum and a maximum threshold have been applied to the peaks in the PSD between 5.5 and 10-$\sigma$. See the text for details.}
\label{virgo}
\end{figure*}

\section{Application to the VIRGO/SPM data}
In the framework of the PHOEBUS collaboration \citep{App2006}, we have applied the same methodology to 3473 days of the sum of the 3 SPM channels of the VIRGO experiment. Unfortunately, no clear structures appear in the PS. This can be a consequence of the different solar background comparing disk-integrated velocity and intensity data.

In order to go a step further, we put to zero all the points above or below a given threshold before computing the PS. Left panel of Fig.~\ref{virgo} shows the maximum of the highest peak between 22 and 26 in the PS when the level of the threshold is continuously increased (or decreased). For example, the maximum is between 2 and $\sim$ 4 when we put to zero all the points above a threshold changing from 2 to 9 (continuous line in the figure). However, when we remove those peaks in the PS with lower power (less than the threshold, dashed line in the figure) we see that for a threshold between 5.5 and 6.5,  we have a peak in the studied region with an amplitude of $\sim 6$. Right panel of Fig.~\ref{virgo} shows the PS computed from a PSD where only peaks with S/N ratios within the range 5.5 to 10 have been considered.  A similar peak-structure to those found using GOLF data can be clearly seen between 22-26 min. The pattern seems to be larger than the GOLF one. As a consequence of the threshold done, only 50 bins in the PSD are responsible for this structure (the PSD of the Virgo data between 25 and 140 $\mu$Hz has more than 32000 points). Thus, a visual identification of these peaks can be directly done in the PSD. 

Further studies should be carried out by the PHOEBUS group before fully understand the VIRGO/SPM data. If the signal is confirmed, it rules out a GOLF instrumental origin. However, as both instruments are on the same spatial platform an external confirmation (i.e. BiSON or GONG data) would be extremely useful. Finally, the deep study of the non disk-integrated MDI/SoHO data would bring the opportunity to access higher degree modes with new valuable information on the inner core.

\section{Conclusions}
In the present work we have shown the presence of periodic signals in the PSD of the GOLF spectrum when it is expressed in period. We have shown by the Monte Carlo simulations that it is very unlikely that these patterns were due to pure noise with 99.49 $\%$ confidence level. Assuming that it is due to g-mode signals, it can only be produced by the presence of $\ell$=1 modes with $-4 \le n\le -26$. A direct comparison with theoretical g modes showed that the $H_0$ hypothesis can be rejected with more than 99.86$\%$ confidence level. 

From this analysis we start to have access to the dynamics of the internal core and we have shown that the core seems to rotate in average between 3 to 5 times faster than the rest of the radiative zone when unrealistic ``rotation-step'' profile is used. More work should be done in the future to introduce more realistic core rotation law profiles and also to use the information corresponding to the higher degree modes. 

Another important consequence of the analysis performed here is that the individual structure of the g-mode multiplets should be more complicated than what, up to now, has been thought. To simulate realistic PSD (low S/N) we need a higher number of bins containing signal. As we have seen, at least, three different scenarios are compatible with a significant reduction of the required S/N.  Further studies will be necessary to discriminate between them and to improve our knowledge of this modes. 

\section*{Acknowledgments}
The GOLF experiment is based upon a consortium of institutes (IAS, CEA/Saclay, Nice and Bordeaux Observatories from France, and IAC from Spain) involving a large number of scientists and engineers, as enumerated in \citep{Gabriel1995}. SoHO is a mission of international cooperation between ESA and NASA. The authors thank T. Appourchaux and W.J. Chaplin for useful discussion and comments and all the PHOEBUS group for providing the VIRGO/SPM data.

\end{document}